\documentclass[aps,prl,twocolumn,groupedaddress]{revtex4-1}

\usepackage{graphicx}
\usepackage{dcolumn} 
\usepackage{bm}      

\begin{document}

\title{Direct observation of the Fermi surface in an ultracold atomic gas}
\author{T. E. Drake, Y. Sagi, R. Paudel, J. T. Stewart, J. P. Gaebler, and D. S. Jin}
\email[Electronic address: ]{jin@jilau1.colorado.edu}
\homepage[URL: ]{http://jilawww.colorado.edu/~jin/}

\affiliation{JILA, National Institute of Standards and Technology and the University of Colorado, and the Department of Physics, University of Colorado, Boulder, CO 80309-0440, USA}

\date{\today} \begin{abstract}

The ideal (i.e. noninteracting), homogeneous Fermi gas, with its characteristic sharp Fermi surface in the momentum distribution, is a fundamental concept relevant to the behavior of many systems.  With trapped Fermi gases of ultracold atoms, one can realize and probe a nearly ideal Fermi gas, however these systems have a nonuniform density due to the confining potential.  We show that the effect of the density variation, which typically washes out any semblance of a Fermi surface step in the momentum distribution, can be mitigated by selectively probing atoms near the center of a trapped gas.  With this approach, we have directly measured a Fermi surface in momentum space for a nearly ideal gas, where the average density and temperature of the probed portion of the gas can be determined from the location and sharpness of the Fermi surface. 

\end{abstract}


\maketitle

The homogeneous Fermi gas is a widely used model in quantum many-body physics and is the starting point for theoretical treatment of interacting Fermi systems.  The momentum distribution for an ideal Fermi gas is given by the Fermi-Dirac distribution:
\begin{equation}\label{eq:FD}
n(k)=\frac{1}{e^{\left(\frac{\hbar^2k^2}{2m}-\mu\right)/k_BT}+1},
\end{equation}
where the $n(k)$ is the average occupation of a state with momentum $\hbar k$, $m$ is the fermion mass, $\mu$ is the chemical potential, $k_B$ is Boltzmann's constant, and $T$ is the temperature.
Surprisingly, to our knowledge, the momentum distribution of an ideal Fermi gas, with its sharp step at the Fermi momentum, $\hbar k_F$, has not been directly observed in experiments.  For the vast majority of Fermi systems, such as electrons in materials, valence electrons in atoms, and protons and/or neutrons in nuclear matter, one always has an interacting system.  A dilute Fermi gas of atoms opens new possibilities with its low density, access to the momentum distribution through time-of-flight imaging, and controllable interparticle interactions. 
However, these trapped gases have nonuniform density, which has prevented the observation of a sharp step in their momentum distribution and, more generally, can complicate comparisons with theory. 

If the change in the trapped gas density is small on the length scale of the relevant physics, one can apply a local density approximation. Measurements can then be compared to theory by integrating the prediction for a homogeneous gas over the density distribution of the trapped gas.  While the agreement between experiment and theory can be quite good, characteristic features such as a sharp Fermi surface in $k$-space can be lost in trap-averaged data. For rf spectroscopy and for thermodynamic measurements, recent work has used $\it{in}$-$\it{situ}$ imaging of trapped gases combined with knowledge of the trapping potential to yield results that can be directly compared to homogeneous Fermi gas theory \cite{Ketterle2008,Nascimbene2010,Horikoshi2010,Ku2011}. However, this technique cannot probe the momentum distribution, which requires a sudden release of the gas from the trap followed by ballistic expansion and imaging.  In this paper, we introduce a method to measure the momentum distribution locally in a trapped Fermi gas and present a direct observation of the Fermi surface in $k$-space.

\begin{figure}
\includegraphics[width=8.0 cm]{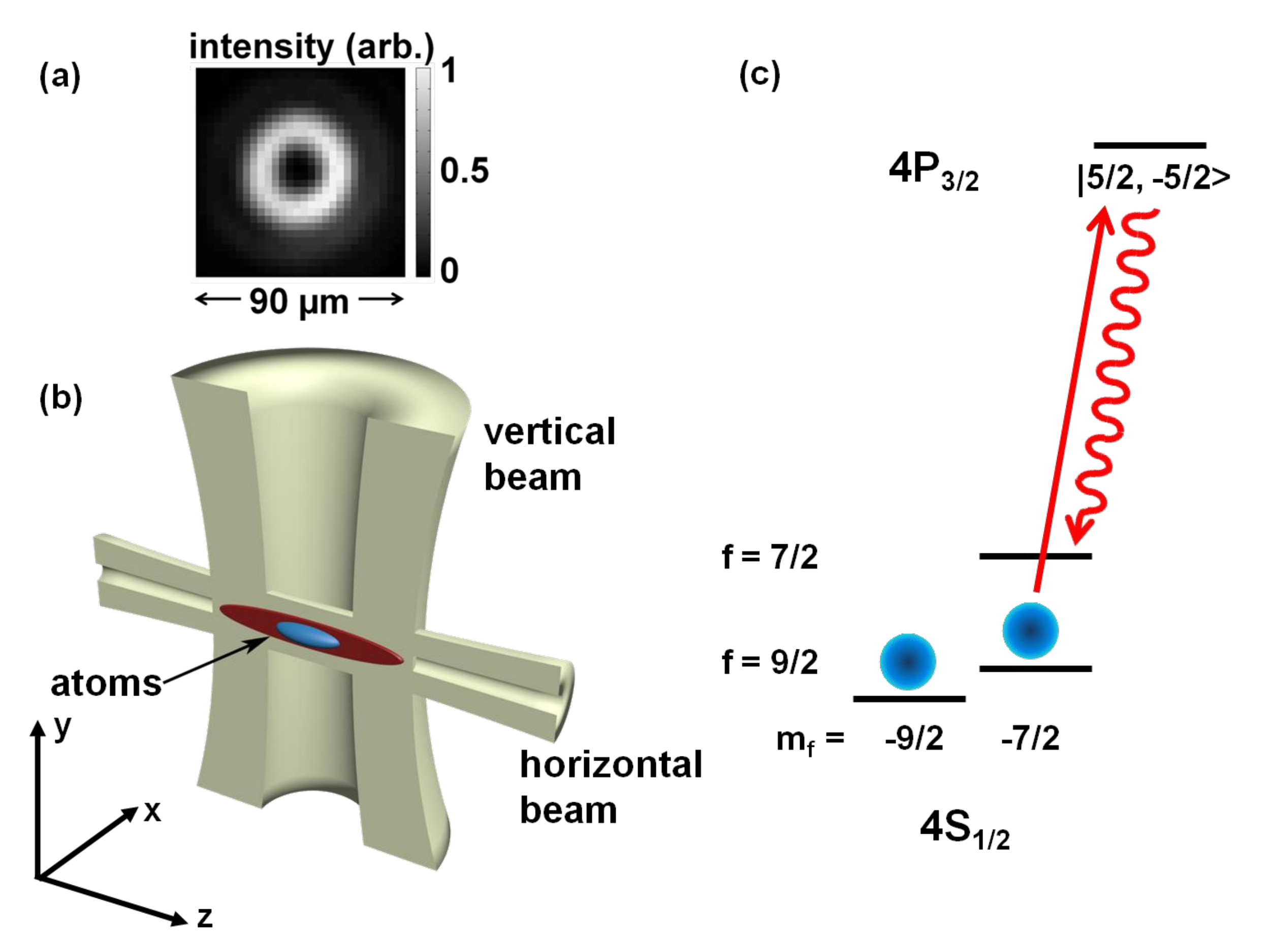}
\caption{\textbf{Spatially selective optical pumping with hollow light beams.} (a) An image of the horizontal hollow light beam.  (b) Illustration of the two intersecting hollow light beams used to optically pump atoms at the edges of the trapped atom cloud. (c) Schematic level diagram showing the optical pumping and imaging transitions. The relevant states at $B=208.2$ G are labeled with the hyperfine quantum numbers of the $B=0$ states to which they adiabatically connect.} \label{donut}
\end{figure}

Previous ultracold gas work has explored box-like confining potentials with relatively sharp ``walls" to create a more homogeneous gas \cite{Raizen2005,Davidson2002}; however, achieving a quantum degenerate gas in these larger volume traps has proven to be challenging. Alternatively, within the local density approximation, one could probe a gas with a more uniform density simply by probing a small fraction of the atoms within the trapped gas.  To obtain spatial selectivity in addition to momentum information, we use optical pumping by shaped light beams \cite{Ketterle1997} to select atoms at the center of the cloud (see Fig. \ref{donut}) as we release the trapped gas for ballistic expansion and imaging. We observe the emergence of a step-like momentum distribution as the fraction of atoms probed decreases.  We find that when probing $40\%$ or less of the atoms, the observed momentum distribution is consistent with that of a homogenous gas, where
the width and position of the Fermi surface reflect the average temperature and density of the probed portion of the gas.

We begin with a quantum degenerate gas of $N=9\times10^4$ $^{40}\rm{K}$ atoms in an equal mixture of the $|f,m_f\rangle= |9/2, -9/2\rangle$ and
$|9/2,-7/2\rangle$ spin states, where $f$ is the quantum number denoting the total atomic spin and $m_f$ is its projection.  The atoms are confined in a cylindrically symmetric, crossed-beam optical trap characterized by a radial trap frequency $\nu_r$ of $214$ Hz and an axial trap frequency $\nu_z$ of $16$ Hz.  To facilitate future application of this technique to probing strongly interacting Fermi gases, we work at a magnetic field that is near a Feshbach resonance between the initial two spin states.  We take data at $B = 208.2$ G where the scattering length $a$ between atoms in the $|9/2,-9/2\rangle$ and $|9/2,-7/2\rangle$ states is approximately $-30 ~a_0$ \cite{Stewart2008}, where $a_0$ is the Bohr radius.  Here, the gas is very weakly interacting, with a dimensionless interaction strength of $k_F a=0.011$.  

Our measurements probe only the $|9/2,-7/2\rangle$ spin component. To selectively probe atoms near the center of the cloud, we take advantage of the many spin states of $^{40}\rm{K}$ and use two intersecting hollow light beams to optically pump atoms into a spin state that is dark to our imaging (see Fig. \ref{donut}b). The hollow light beams are resonant with the transition from the $|9/2,-7/2\rangle$ state to the electronically excited $|5/2,-5/2\rangle$ state (see Fig. \ref{donut}c).  Atoms in this excited state decay by spontaneous emission with a branching ratio of $0.955$ to the $|7/2,-7/2\rangle$
ground state and $0.044$ to the original $|9/2,-7/2\rangle$ state.

The beams have a Laguerre-Gaussian spatial mode with an angular index of 2 (see Fig. \ref{donut}a) and are formed using an absorption mask patterned with a forked diffraction grating with two dislocations \cite{Heckenberg1992}. At the focus, the light intensity is given by
\begin{equation}
I(r)=\frac{P}{\pi w^2} \left(\frac{2r^2}{w^2}\right)^{2} e^{-\frac{2r^2}{w^2}}
\end{equation}
where $P$ is the total optical power, $r$ is the radial coordinate transverse to the direction of beam propagation, and $w$ is the waist.
The first beam propagates along the vertical ($y$) direction, is linearly polarized, and has a waist of 186 $\mu$m.  Given the elongated shape of the trapped gas, this beam is primarily spatially selective along the long axis ($z$) of the cloud. The second beam propagates along the axial ($z$) direction of the cylindrically symmetric trap, parallel to $B$, and is circularly polarized.  This beam has a waist of 16.8 $\mu$m and selectively optically pumps atoms based on their location along $x$ and $y$.  The beams are independently aligned onto the atom cloud by looking at the fraction of atoms probed and minimizing shifts in the position and center-of-mass momentum of the probed atoms after 10 ms of expansion.

To probe the momentum distribution of the central part of the trapped gas, we first turn off the trap suddenly and illuminate the atoms with the vertical hollow light beam, followed immediately by pulsing on the horizontal beam.  The power in the beams in on the order of 10s to 100s of nW and is varied to control the fraction of atoms that are optically pumped out of the $|9/2,-7/2\rangle$ state. Each beam is pulsed on for 10 to 40 $\mu$s, with the pulse durations chosen such that the fraction of atoms optically pumped by each of the two beams is roughly equal (within a factor of two).
We then image the remaining atoms in the $|9/2,-7/2\rangle$ state after 10 ms or 12 ms time of flight \cite{imaging}.  The imaging light propagates along the $z$ direction and we apply an inverse Abel transform to the 2D image (assuming spherical symmetry in $k$-space) to obtain the 3D momentum distribution, $n(k)$.

\begin{figure}
\includegraphics[width= 8 cm]{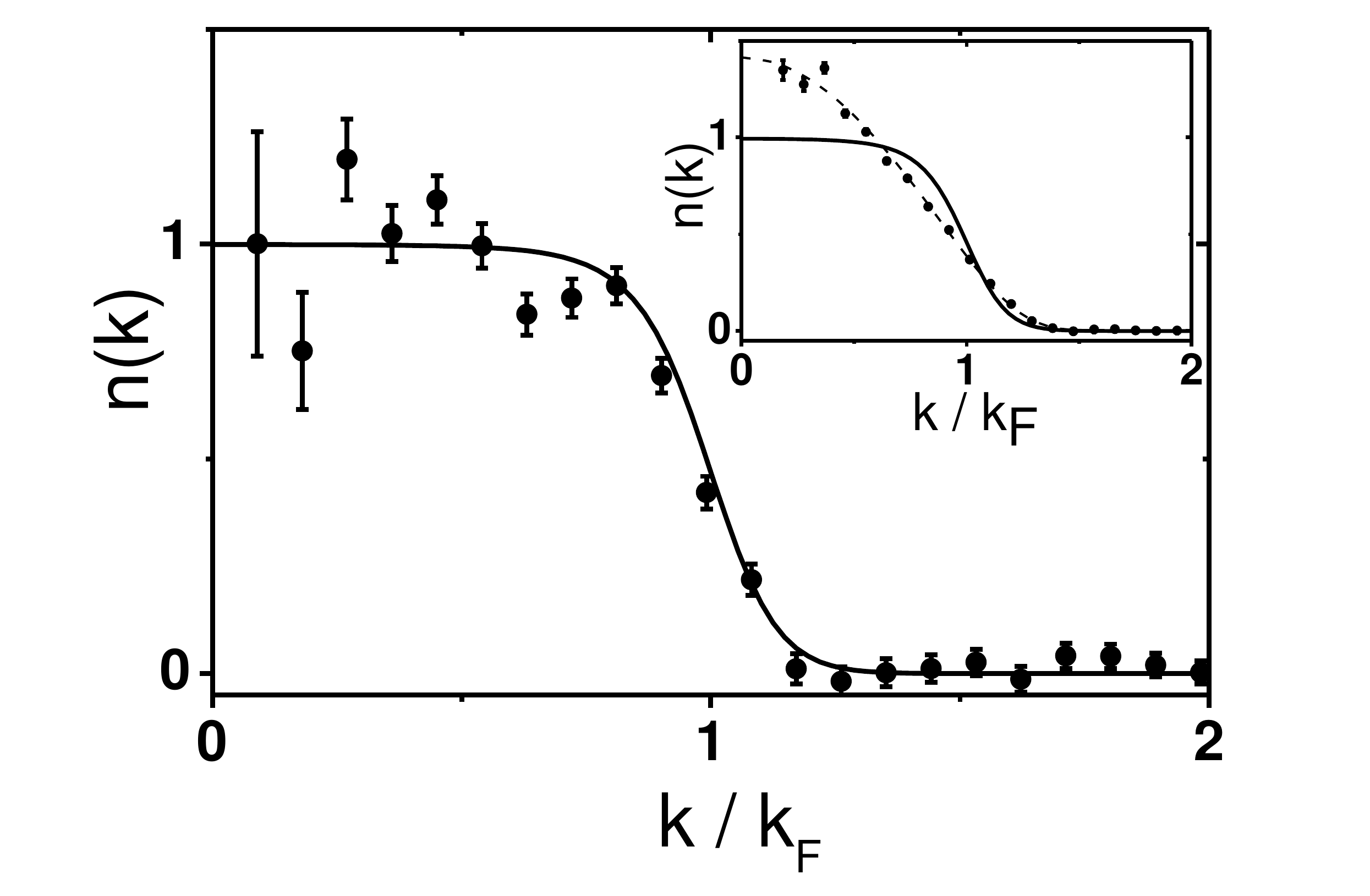}
\caption{\textbf{Measured momentum distribution for a weakly interacting Fermi gas.}  The momentum distribution of the central 16$\%$ of a harmonically trapped gas is obtained from the average of twelve images.  The solid line shows a fit to a homogeneous momentum distribution, where $T$ is fixed to the value obtained from the trap-averaged distribution and $k_F$ is the only fit parameter. (Inset) For comparison, we show the trap-averaged distribution, taken from an average of six images, with fits to the expected momentum distribution for an ideal gas in a harmonic trap (dashed line) and to the homogeneous momentum distribution (solid line).  }
\label{nk}
\end{figure}

In Fig. \ref{nk}, we show normalized momentum distributions measured with and without using the hollow light beams.   As was first seen in Ref. \cite{DeMarco2001}, the trap-averaged momentum distribution for the Fermi gas is only modestly distorted from the Gaussian distribution of a classical gas.  The dashed line in the inset to Fig. \ref{nk} shows a fit to the expected momentum distribution for a harmonically trapped ideal Fermi gas, from which we determine the temperature of the gas to be $T/T_{F,\rm{trap}} = 0.12 \pm 0.02$.
Here, the Fermi temperature for the trapped gas is given by $T_{F,\rm{trap}}=E_{F,\rm{trap}}/k_B$, where $E_{F,\rm{trap}}=h(\nu_r^2 \nu_z)^{1/3} (6 N)^{1/3}$ is the Fermi energy for the trapped gas. 
After optical pumping with the hollow light beams so that we probe the central $16\%$ of the atoms, the measured momentum distribution (main part of Fig. \ref{nk}) has a clear step, as expected for a homogeneous Fermi gas described by Eqn. 1.  

For a sufficiently small density inhomogeneity, the momentum distribution should look like that for a homogeneous gas at some average density.  
To characterize this, we fit the normalized distributions to the prediction for an ideal homogeneous gas (solid lines).  The homogeneous gas distribution is described by its temperature and density.  We fix $T$ to that measured for the trapped gas, which leaves only a single fit parameter, $k_F$, that characterizes the density.  The momentum distributions are then plotted as a function of the usual dimensionless momentum, $k/k_F$.  The momentum distribution of the central $16\%$ of the trapped gas fits well to the homogeneous gas result, while the trap-averaged momentum distribution clearly does not.

\begin{figure}
\includegraphics[width= 8 cm]{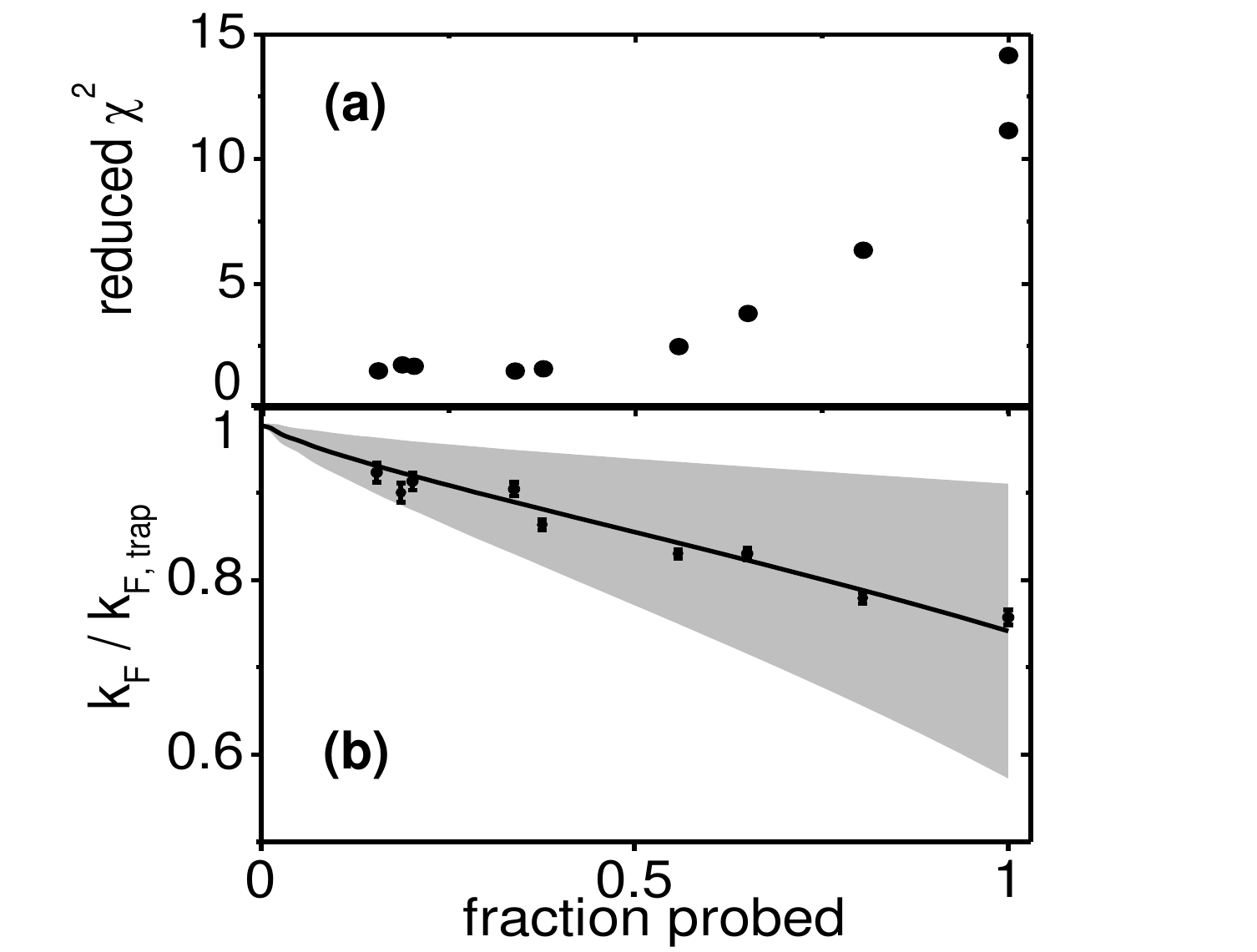}
\caption{\textbf{Fit results as a function of the fraction of atoms probed.} (a) As the fraction of atoms probed is decreased (by increasing the power in the hollow light beams), the reduced $\chi^2$ for the fit to a homogeneous gas momentum distribution decreases dramatically and approaches a value 1.6 for fractions less than $40\%$. (b) The best fit value $k_F$ increases as we selectively probe fewer atoms from the highest density part of the cloud.  A model of the optical pumping by the hollow light beams yields an average local $k_F$ indicated by the solid line. As an indication of the density inhomogeneity, the shaded region shows the spread (standard deviation) in $k_F$ from the model.} 
\label{kF}
\end{figure}

In order to quantify how well the measured momentum distribution is described by that of a homogeneous gas, we look at the reduced $\chi^2$ statistic in Fig. \ref{kF}a.  The reduced $\chi^2$ is much larger than 1, indicating a poor fit, for the trap-averaged data due the fact that the density inhomogeneity washes out the Fermi surface. As we probe a decreasing fraction of atoms near the center of the trap, $\chi^2$ decreases dramatically and approaches a value of 1.6 for fractions smaller than $40\%$. 

The single fit parameter $k_F$ characterizes the density of the probed gas and should increase as we probe fewer atoms near the center of the trap. Fig. \ref{kF}b displays the fit value $k_F$, in units of $k_{F,\rm{trap}}=\sqrt{2mE_{F,\rm{trap}}}/\hbar$.  As expected, $k_F$ increases as the fraction of atoms probed decreases.  We have developed a model of the spatially selective optical pumping by the hollow light beam, which we discuss below.  The model result for the average local $k_F$, $\left< k_F \right>$, of the probed gas is shown with the solid line in Fig. \ref{kF}b, and we find that this agrees well with the fit $k_F$, even when the measured momentum distributions clearly do not look like that of a homogeneous gas. Using the model, we calculate the variance $\delta^2$ of the local $k_F$, and the shaded region in Fig. \ref{kF}b shows $\left< k_F \right> \pm \delta$. In the region where the reduced $\chi^2$ indicates that the measured $n(k)$ fits well that for a homogeneous gas (fraction probed $<40\%$), $\delta/\left< k_F \right> < 0.08$.

\begin{figure}
\includegraphics[width= 8 cm]{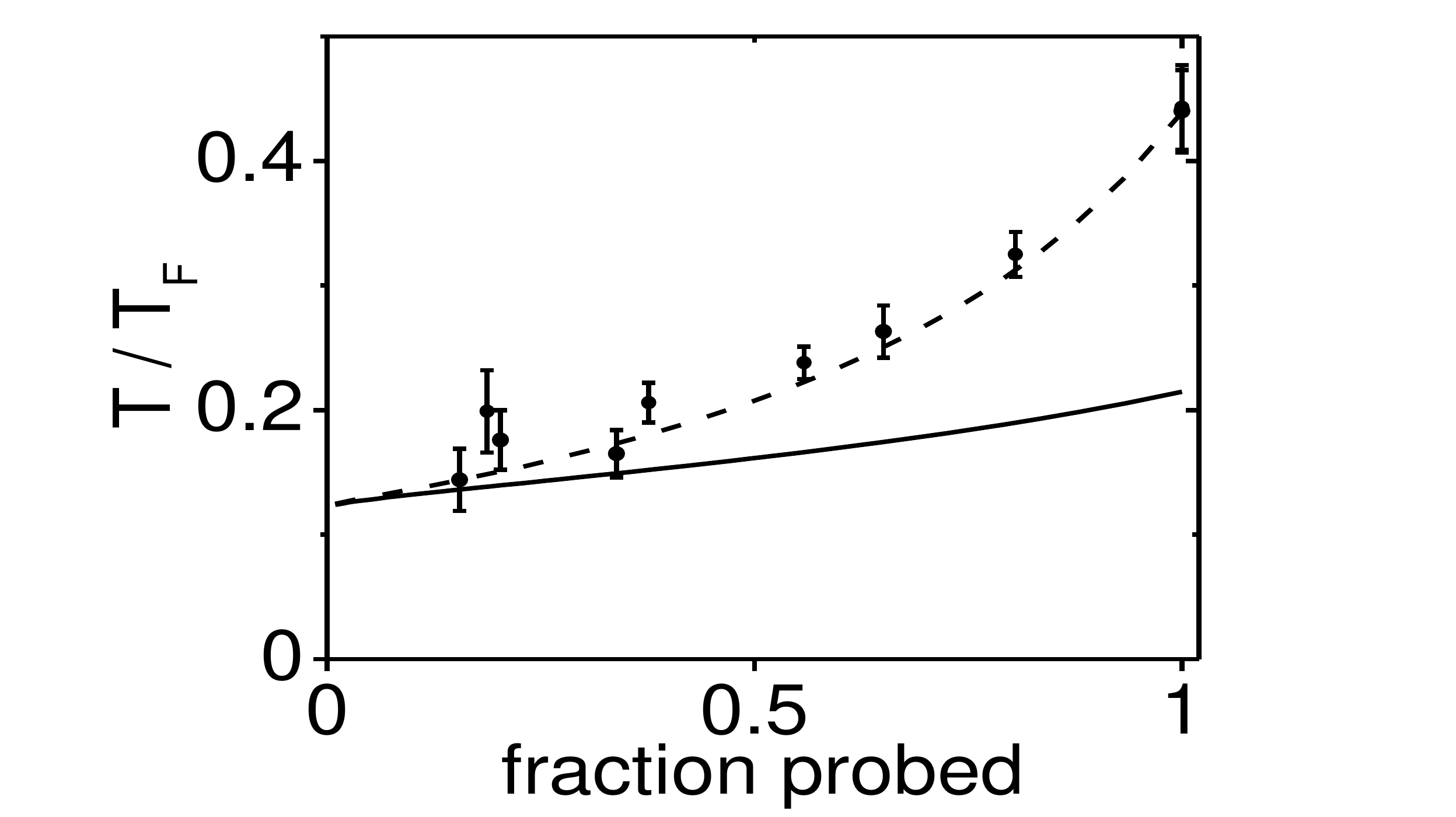}
\caption{\textbf{Measured $T/T_F$ vs the fraction of atoms probed.} Here, we fit the measured momentum distribution to a homogeneous gas distribution with two free parameters, $T/T_F$ and $k_F$. The density inhomogeneity of the probed gas results a $T/T_F$ that is much larger than that expected from the calculated average density of the probed gas (solid line). A sharp Fermi surface, characterized by a small fit $T/T_F$, emerges as the fraction of atoms probed decreases.   The dashed line shows the result of fitting to model calculations of the probed momentum distribution, which agrees well with the data.} \label{ttf}
\end{figure}

Instead of fixing $T$ to the value obtained from fitting the trapped gas momentum distribution, we can also look at measuring the temperature by fitting to a homogeneous gas distribution, where both $k_F$ and $T/T_F$ are fit parameters.  In this case, a large density inhomogeneity that washes out the Fermi surface will result in an artificially high fit value for $T/T_F$.  This can be seen in Fig. \ref{ttf}.  For comparison to the data, the solid line shows the average $T/ \left< T_F \right>$ for the probed gas calculated using our model. Here, $T$ is fixed and the dependence on the fraction probed comes from the fact that the average density, and therefore the average local $T_F$, increases as we probe a smaller fraction of atoms that were near the center of the trapped gas.  The fit $T/T_F$ approaches the average value from the model as we reduce the fraction of atoms probed, and for $<40\%$ probed, the two are consistent within our measurement uncertainty.  For the smallest fraction probed (data shown in Fig. \ref{nk}), the best fit value is $T/T_F = 0.14 \pm 0.02$.  As a check of the model, we can also calculate $n(k)$ for the probed gas and fit this to the homogeneous gas distribution; the results (dashed line in Fig. \ref{ttf}) agree well with the data. 

In modeling the effect of optical pumping with the hollow light beams, we assume that only atoms that do not scatter a photon are probed. The probability to scatter zero photons from each beam is taken to be $P_{i} = \exp \! \left( {- \gamma_i \tau_i \sigma} \right)$, where
$\tau_i$ is the pulse duration and the subscripts $i=1,2$ denote the two hollow light beams. The photon flux is given by $\gamma_i={I_i \lambda}/({h c})$, where $I_i$ is the position-dependent intensity, $c$ is the speed of light, and $\lambda=766.7$ nm is the wavelength. For the optical absorption cross section, we use $\sigma={3 \lambda^2\eta}/({2 \pi})$, where $\eta=0.044$ is the branching ratio back to the initial state. For the intensities, we use Eqn. 2 and make the approximation that $w$ is constant across the cloud.

\begin{figure}
\includegraphics[width= 8.0 cm]{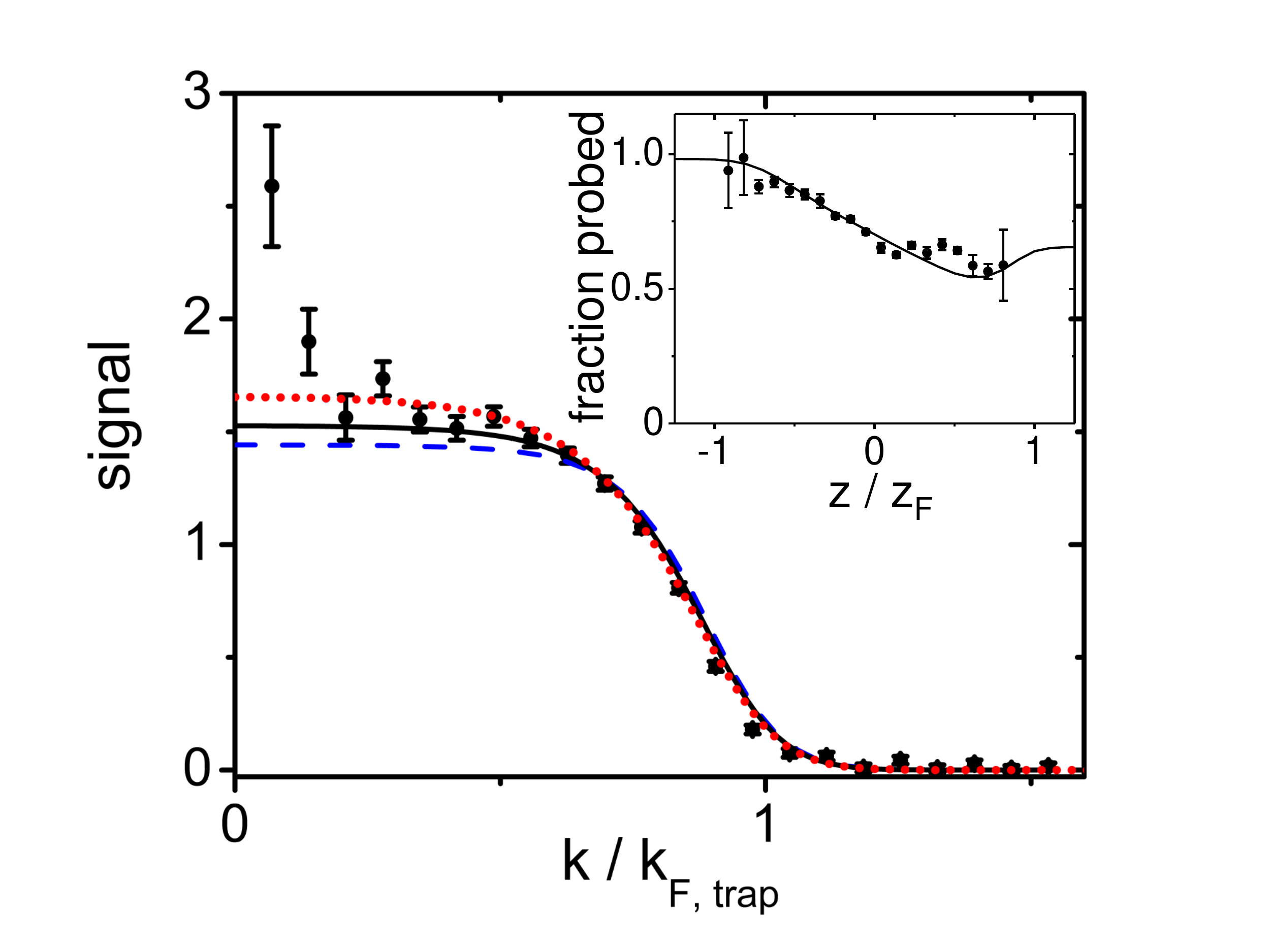}
\caption{\textbf{Modeling the spatially selective optical pumping.} We compare the normalized momentum distribution of the central $38\%$ of the atoms to three different models (dotted, solid, and dashed lines; see text). The data (circles) are obtained from an average of four images.  We find that the attenuation of the hollow light beams (inset) does not strongly affect the predicted final momentum distribution when probing a small fraction of the gas. (Inset) We take images of the cloud after a short (1.3 ms) expansion and compare data with the horizontal hollow light beam and without any optical pumping in order to measure the fraction of atoms probed (circles) vs $z/z_F$, where $z_F=\sqrt{\frac{2 E_{F,\rm{trap}}}{m(2\pi \nu_z)^2}}$.  For this data, the fraction probed is 71$\%$.  The prediction of our model (solid line), which includes attenuation of the hollow light beam as it propagates through the cloud, agrees well with the data.} \label{atten}
\end{figure}

Attenuation of the hollow light beam as it propagates through the atom cloud is observable in the long direction of the cloud (along $z$), as seen in the inset of Fig. \ref{atten}. To include this effect, we consider the two hollow light beam pulses sequentially, and we assume that the number of photons absorbed locally equals the number of optically pumped atoms.  Interestingly, the model predicts that the attenuation results in a smaller density variance in the probed gas when compared to a model that ignores attenuation but where we adjust the beam powers to probe the same fraction of the atoms.  This effect is relatively small and decreases as one probes a smaller fraction of the gas.  This can be seen in Fig. \ref{atten} where we show the measured momentum distribution for the central $38\%$ of the atoms compared to three different models, each of which is adjusted to give the same probed fraction.  The solid line is the model explained above, which includes attenuation, while the dotted line show the result when we ignore the depletion of the hollow light beams.  For comparison, the dashed line shows the expected distribution if one selects atoms in a cylindrical volume with sharp boundaries.

In conclusion, we have directly observed the Fermi surface in the momentum distribution of a weakly interacting Fermi gas.  To do this, we probe the central region of a harmonically trapped gas. A concern with this approach is that one might be left with very little signal after selecting a small enough region to approximate a homogeneous gas. However, for a gas at $T/T_{F,\rm{trap}}=0.12$ and our typical measurement precision, we find that probing the central $40\%$ (or less) of the gas is sufficient to approximate a homogeneous gas. 

 We anticipate that this ability to obtain local momentum distributions for a trapped gas can be applied to probe interacting Fermi gases, including strongly interacting gases in the regime of the BCS-BEC crossover \cite{Regal2007}.  Examples of probes of these systems that require momentum resolution and could benefit from the removal of the effects of density inhomogeneity include measurements of the condensate fraction \cite{Regal2004,Zwierlein2004}, determination of the contact parameter \cite{Tan2008a,Tan2008b,Tan2008c} from the tail of the momentum distribution \cite{Stewart2010}, and measurements of the Fermi spectral function using atom photoemission spectroscopy \cite{Stewart2008}.  Measurements of $n(k)$ for a homogeneous Fermi gas with interactions could also be used to test theoretical predictions, such as the quasiparticle weight for a Fermi liquid \cite{Landau}.  More generally, this technique should be broadly applicable and could provide access to the lowest entropy part of a trapped gas of ultracold bosons, fermions, or mixtures, for a variety of different trapping potentials and interaction strengths.

\begin{acknowledgments}
We acknowledge funding from the NSF and NIST.  We thank B. DeMarco and J. Zirbel for initial assistance with the hollow light beams.
\end{acknowledgments}


\begin{thebibliography}{10}
\bibitem{imaging}
To optimize the signal-to-noise ratio in the image, we use an rf $\pi$-pulse to
  transfer the remaining $|9/2,-7/2\rangle$ atoms to the $|9/2,-9/2\rangle$
  state and then image with light resonant with the cycling transition. Just
  prior to this rf pulse, we clean out the $|9/2,-9/2\rangle$ state by
  optically pumping these atoms to the upper hyperfine ground state ($f$=7/2).
\end{thebibliography}
\end{document}